\documentclass[journal]{IEEEtran}

\usepackage{amsmath,amsfonts,amssymb}
\usepackage{array}
\usepackage{textcomp}
\usepackage{stfloats}
\usepackage{url}
\usepackage{graphicx}
\usepackage{cite}
\usepackage[utf8]{inputenc}
\usepackage[T1]{fontenc}
\usepackage[dvipsnames,table]{xcolor}
\usepackage{soul}
\sethlcolor{yellow}

\newif\ifshowchg
\showchgtrue
\ifshowchg

\else

\fi

\graphicspath{{figs/}}

\begin{document}

\title{TLS and Quasiparticle Loss in Thin-Film Aluminum CPW Resonators:
A Modified Model and Design Implications}

\author{Carolyn~G.~Volpert,~\IEEEmembership{Graduate~Student~Member,~IEEE,}
        Emily~M.~Barrentine,
        Alberto~D.~Bolatto,
        Ari~Brown,
        Jake~A.~Connors,
        Thomas~Essinger-Hileman,
        Larry~A.~Hess,
        Vilem~Mikula,
        Thomas~R.~Stevenson,
        and~Eric~R.~Switzer%
\thanks{This work was supported by the NASA ROSES APRA Program via
Grant NNH21ZDA001N-APRA and Grant APRA 1263 17-APRA17-0077.}
        
\thanks{C.~G.~Volpert is with the Department of Astronomy,
        University of Maryland, College Park, MD~20742, USA,
        and also with NASA Goddard Space Flight Center,
        Greenbelt, MD~20771, USA
        (e-mail: cvolpert@astro.umd.edu).}%
\thanks{E.~M.~Barrentine, A.~Brown, J.~A.~Connors,
        T.~Essinger-Hileman, L.~A.~Hess, V.~Mikula,
        T.~R.~Stevenson, and E.~R.~Switzer are with
        NASA Goddard Space Flight Center,
        Greenbelt, MD~20771, USA.}%
\thanks{A.~D.~Bolatto is with the Department of Astronomy,
        University of Maryland, College Park, MD~20742, USA.}%
\thanks{Manuscript submitted April 2025.}}

\markboth{IEEE Transactions on Applied Superconductivity,~Vol.~XX,
          No.~X, XXXX~2025}%
{Volpert \MakeLowercase{\textit{et al.}}: TLS and Quasiparticle Loss in
 Thin-Film Al CPW Resonators}

\maketitle

\begin{abstract}
As superconducting kinetic inductance detectors (KIDs) continue to grow
in popularity for sensitive submillimeter detection and other applications,
there is a drive to advance toward lower-loss devices. We present
measurements of diagnostic thin-film aluminum coplanar waveguide (CPW)
resonators designed to inform ongoing KID development at NASA Goddard
Space Flight Center. The resonance frequencies span
$f_0 = 3.5$--$4\,\mathrm{GHz}$ and include quarter-wave and half-wave
resonators with varying coupling capacitor designs. We present
measurements of the device film properties and an analysis of the dominant
mechanisms of loss in the resonators measured in a dark environment,
demonstrating quality factors of
$Q_i^{-1} \approx 3.64$--$8.57 \times 10^{-8}$.
We observe an enhanced level of suppression in the loss contributions from
two-level systems (TLS) at intermediate-to-high read powers, and a regime
at these powers and low temperatures where contributions from intrinsic
processes $Q_{i,\mathrm{other}}^{-1}$ dominate the total loss. We also
observe deviations from the standard TLS loss model at low powers and
temperatures below $60\,\mathrm{mK}$, and use a modified model to
describe this behavior.
\end{abstract}


\section{Introduction}
\IEEEPARstart{S}{uperconducting} microwave resonators have gained traction
in recent years both for applications in optical detection---particularly at
visible, far-infrared (FIR), and sub-millimeter (sub-mm) frequencies---and
as components in quantum computing circuits. These resonators are attractive
for use in astronomical instruments as kinetic inductance detectors (KIDs)
due to their high sensitivity paired with the unique advantage of
frequency-domain multiplexing.

One present challenge facing applied superconducting resonator technology
is the need to mitigate and otherwise predictably model sources of signal
loss, both through optimizing fabrication processes and establishing
dynamic measurement-based models. The relative novelty and diversity of
applications for these devices drives the field to seek more ideal
materials, fabrication processes, and integrated designs to push devices
towards fundamental performance limits. Both KID and quantum computing
applications incentivize scaling up resonator array sizes while maintaining
high sensitivity, low noise, and array uniformity. This places special
importance upon understanding the loss mechanisms of superconducting
microwave resonators, which manifest as decoherence in quantum circuits and
as reduced optical responsivity and increased noise in optical detectors.

The optical responsivity of a material is determined by its internal
quality factor $Q_i$ and its kinetic inductance fraction $\alpha$. Materials
with both high $Q_i$ and high $\alpha$ provide greater photon sensitivity.
However, imperfections in the fabrication process can induce loss by
introducing parasitic two-level systems (TLS). This motivates choosing
materials and fabrication techniques that prevent contaminants when possible,
and selecting for contaminants that generate fewer TLS when necessary. For
KIDs it is also important to select a material's superconducting energy gap
$\Delta_0$ such that $h\nu < 2\Delta_0$, constraining KID materials to match
the energy bands of the photons being detected. Considering this, TiN, NbTiN,
Nb, Re, and Al have arisen as popular materials in infrared and sub-mm KID
fabrication~\cite{mcrae2020materials}. Here we present measurement results
for thin-film Al on Si CPW resonators designed for far-infrared astronomy,
focusing on intrinsic film properties and sources of loss.

\section{Experimental Setup}

\subsection{Device Design and Fabrication}

The device contains sixteen CPW resonators coupled to a central CPW
feedline. The resonators were wet-etched into a 23\,nm thick film of Al
sputter-deposited on Si. The single-layer device architecture was chosen to
simplify fabrication, provide rapid turnaround of test devices, and to be
broadly applicable to measuring the properties of several superconducting
materials. The device cycles through finger coupling-capacitor structures of
three lengths ($0\,\mu\mathrm{m}$, $10\,\mu\mathrm{m}$, and
$30\,\mu\mathrm{m}$) to provide coverage over a range of internal quality
factors. The feedline was designed with an impedance of $50\,\Omega$ in the
limit of a thick superconducting film (perfect conductor). Fifteen resonators
are $\lambda/4$, and one is $\lambda/2$, coupling to the feedline with a
$30\,\mu\mathrm{m}$ finger-capacitor structure. The device has no optical
inputs and is packaged in a light-tight container; our fundamental resonator
design is intended for antenna-coupled sub-mm detectors. We present data from
a representative $\lambda/4$ resonator and the $\lambda/2$ resonator. The
devices were fabricated in the Detector Development Lab at NASA Goddard Space
Flight Center (GSFC) and all testing was performed at GSFC.

\begin{figure}[!t]
\centering
\includegraphics[width=\columnwidth]{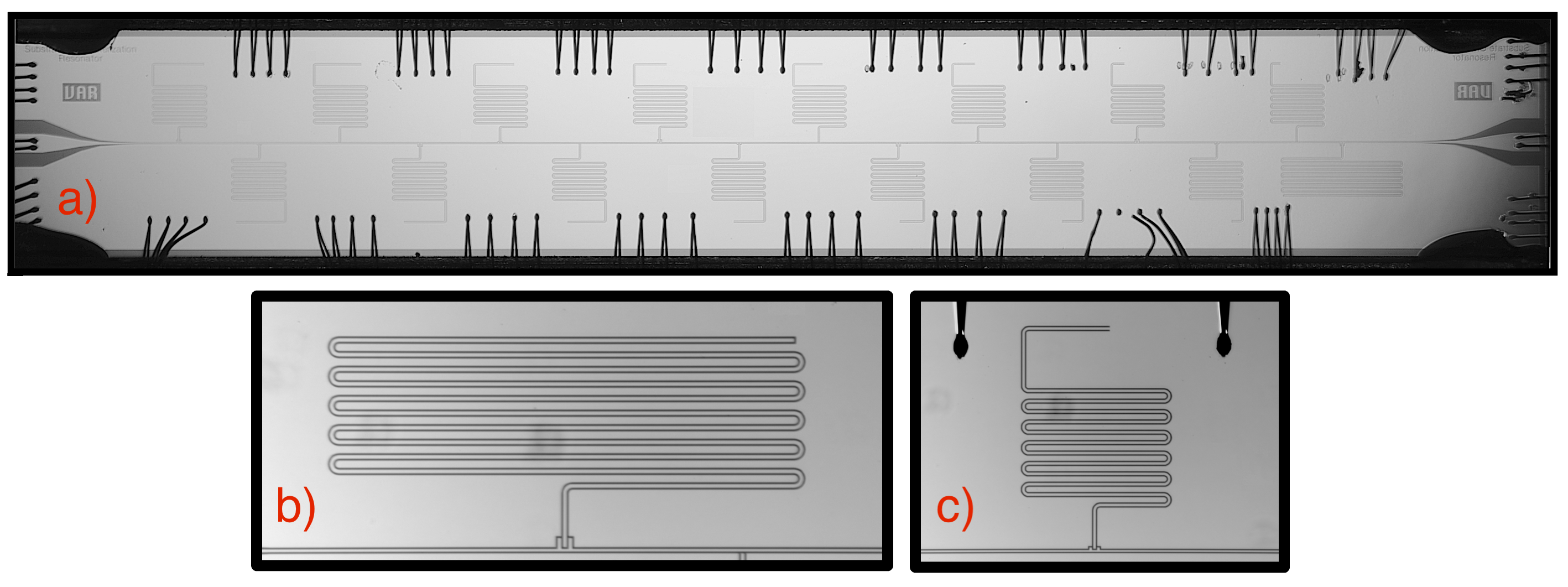}
\caption{(a)~Test chip containing sixteen resonators coupled to a central
feedline. (b)~The $\lambda/2$ resonator. (c)~The $\lambda/4$ resonator.}
\label{fig:chip_photo}
\end{figure}

The fabrication process included standard deposition and patterning steps
designed to mitigate native oxide formation. First, native oxides on the
float-zone Si wafer substrate were cleaned via an HF:H\textsubscript{2}O
(1:10) dip immediately prior to loading into the deposition chamber. Before
deposition, we performed an additional \textit{in situ} reverse-bias clean.
Without this step, we have observed residual irregular patches of Al on the
silicon after etching~\cite{mirzaei2020mu}. We then applied
hexamethyldisilazane (HMDS) and photoresist to the Al for lithographic
patterning and used E6 etchant to wet-etch the microwave circuit. The
photoresist was removed with solvents, and the HMDS was removed with a short
O$_2$ ashing before dicing the wafer into chips.

\begin{table}[!t]
\caption{Device and Film Parameters. Origin: (M)~measured on test device
or pre-dicing wafer, (D)~design value.}
\label{tab:table1}
\begin{center}
\renewcommand{\arraystretch}{1.2}
\begin{tabular}{lll}
\hline
\textbf{Parameter} & \textbf{$\lambda/2$} & \textbf{$\lambda/4$} \\
\hline
Al Thickness (M)                         & \multicolumn{2}{c}{$23\pm1$\,nm} \\
Feedline $T_c$ (M)                       & \multicolumn{2}{c}{$1.31\pm0.03$\,K} \\
Resistivity $\rho_n$ (M)$^\mathrm{a}$   & \multicolumn{2}{c}{$1\times10^{-8}\,\Omega\cdot\mathrm{m}$} \\
Sheet Resistance $R_s$ (M)$^\mathrm{a}$ & \multicolumn{2}{c}{$0.482\,\Omega/\square$} \\
RRR (M)$^\mathrm{a}$                     & \multicolumn{2}{c}{2.650} \\
CPW Width (D)                            & \multicolumn{2}{c}{$10\,\mu\mathrm{m}$} \\
CPW Gap (D)                              & \multicolumn{2}{c}{$5\,\mu\mathrm{m}$} \\
Resonant Frequency $f_r$ (M)             & $3.911$\,GHz & $3.655$\,GHz \\
Resonator Length (D)                     & $7.283$\,mm  & $13.939$\,mm \\
Coupling Length (D)$^\mathrm{b}$         & $30\,\mu\mathrm{m}$ & $10\,\mu\mathrm{m}$ \\
$Q_c\ (\times10^5)$ (M)                  & $0.87$ & $2.58$ \\
Min.\ $Q_i^{-1}\ (\times10^{-8})$ (M)   & $3.64\pm0.11^\mathrm{c}$ & $8.57\pm0.17^\mathrm{d}$ \\
\hline
\multicolumn{3}{l}{$^\mathrm{a}$Measured at 4 K.}\\
\multicolumn{3}{l}{$^\mathrm{b}$Finger length of coupling capacitor.}\\
\multicolumn{3}{l}{$^\mathrm{c}$At $60\,\mathrm{mK}$,
  $\bar{n}_\mathrm{ph}\approx4.8\times10^7$.}\\
\multicolumn{3}{l}{$^\mathrm{d}$At $18\,\mathrm{mK}$,
  $\bar{n}_\mathrm{ph}\approx5.6\times10^6$.}\\
\hline
\end{tabular}
\end{center}
\end{table}

\subsection{Measurement Setup}

\begin{figure}[!t]
\centering
\includegraphics[width=\columnwidth]{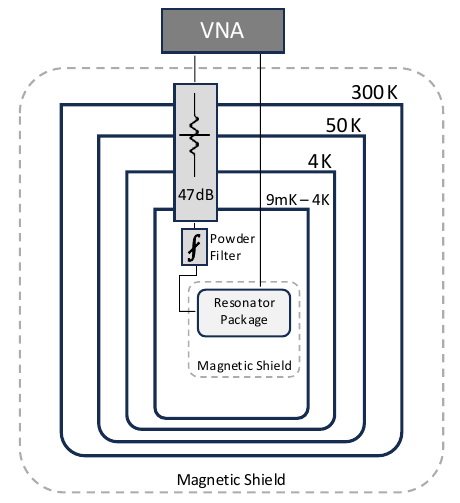}
\caption{Cryogenic testbed configuration. The test chip was situated within
two layers of magnetic shielding on a temperature-controlled cold stage.}
\label{fig:testbed}
\end{figure}

We mounted the device on the temperature-controlled stage of a dilution
refrigerator beneath two layers of magnetic shielding
(Fig.~\ref{fig:testbed}). We read out the device using an Agilent
Technologies PNA-X vector network analyzer (VNA). The read-tone from the VNA
was cumulatively attenuated by $-47\,\mathrm{dB}$ (via microwave cabling and
attenuators) and fed through absorptive thermal blocking powder
filters~\cite{wollack2014impedance} before reaching the device input on the
cold stage. We measured $S_{21}^\mathrm{meas}(f_\mathrm{read})$ while varying
the input tone power ($P_\mathrm{read} = {-137}\,\text{--}\,{-67}\,\mathrm{dBm}$
post-attenuation) and the device temperature
($T_\mathrm{bath} = 9\,\text{--}\,400\,\mathrm{mK}$).

\section{Theory and Analysis}

\noindent We quantify resonator performance using the quality factor $Q_r$,
with $Q_r^{-1} = Q_i^{-1} + Q_c^{-1}$, where $Q_i$ is the internal quality
factor (energy stored per unit of energy lost internally per cycle), and
$Q_c$ is the coupling quality factor (energy stored per unit of energy lost
to coupling).

\subsection{Fitting Transmission Data}

To recover $Q_i$ and each resonator's resonant frequency $f_r$, we fit our
$S_{21}(f_\mathrm{read})$ data using a hanger-type resonator model, which
treats each resonator as a short-circuited transmission line side-coupled to
the central CPW feedline (Fig.~\ref{fig:chip_photo}). We use the diameter
correction method (DCM) version of this model to produce an asymmetric
Lorentzian with a dip in magnitude and a phase shift at $f_r$. The ideal DCM
transmission is~\cite{khalil2012analysis}

\begin{equation}
    S_{21}(x) = \frac{Q_i^{-1} + 2i(x + \delta x)}{Q_r^{-1} + 2ix}
    \label{eq:S21_DCM}
\end{equation}

\noindent where $x(f_\mathrm{read}) = (f_\mathrm{read} - f_r)/f_r$, and
$\delta x$ accounts for the frequency shift arising from impedance mismatches
rotating the resonator response in the complex plane.

We introduce an additional term to account for finite-length feedlines,
internal reflections, and other sources of transmission asymmetry, giving a
total fit of

\begin{equation}
    S_{21}^\mathrm{meas} = g(f_\mathrm{read})\,
                            S_{21}^\mathrm{res}\,
                            e^{i\phi(f_\mathrm{read})}
    \label{eq:S21_total}
\end{equation}

\noindent where

\begin{align}
    g(f_\mathrm{read}) &= g_0 + g_1 x + g_2 x^2 \nonumber \\
    \phi(f_\mathrm{read}) &= \phi_0 + \phi_1 x + \phi_2 x^2
\end{align}

\noindent are a frequency-dependent complex gain factor and phase shift
factor, respectively~\cite{gao2008physics, khalil2012analysis,
probst2015efficient}. We modified the Python package \texttt{scraps} to fit
our large dataset to this model~\cite{carter2016scraps}.

\subsection{Loss Mechanisms}

$Q_i^{-1}$ is a loss metric for a resonator, with smaller values indicating
less transmission loss and greater detector sensitivity. When a photon of
sufficient energy strikes a superconducting film it breaks Cooper pairs and
generates quasiparticles, changing the resonator kinetic inductance. Other
mechanisms can also impact the quasiparticle population, including thermal
generation, read-tone-induced pair breaking, two-level system effects,
radiation loss, magnetic vortices, and stray-light-induced pair breaking.
This work considers the latter three contributions together as a
temperature- and read-power-independent term $Q_{i,\mathrm{other}}^{-1}$.
Without an optical contribution, we model the total resonator loss
as~\cite{gao2008experimental}

\begin{equation}
    Q_i^{-1} = Q_{i,\mathrm{qp}}^{-1}(T_\mathrm{bath}, P_\mathrm{read})
             + Q_{i,\mathrm{TLS}}^{-1}(T_\mathrm{bath}, P_\mathrm{read})
             + Q_{i,\mathrm{other}}^{-1}
    \label{eq:total_loss}
\end{equation}

\noindent where $Q_{i,\mathrm{qp}}^{-1}$ is the contribution from excess
quasiparticles and $Q_{i,\mathrm{TLS}}^{-1}$ is the TLS contribution.

\subsubsection{Quasiparticle Loss}

$Q_{i,\mathrm{qp}}^{-1}$ depends on the number of quasiparticles $N_\mathrm{qp}$
in the resonator~\cite{zmuidzinas2012superconducting}:

\begin{equation}
    Q_{i,\mathrm{qp}}^{-1} = N_\mathrm{qp}
        \cdot \frac{2\alpha}{4 N_0 \Delta V}\, S_1
    \label{eq:Q_qp}
\end{equation}

\noindent where $\alpha$ is the kinetic inductance fraction,
$N_0 = 1.74\times10^{10}\,\mathrm{eV}^{-1}\mu\mathrm{m}^{-3}$ is the
single-spin density of states at the Fermi level for
aluminum~\cite{goldie2012non}, $\Delta = 1.764\,k_B T_c$ is the
superconducting gap energy, $V$ is the resonator volume, and $S_1$ is the
dimensionless Mattis-Bardeen response function~\cite{zmuidzinas2012superconducting}

\begin{equation}
    S_1 = \frac{2}{\pi}
          \sqrt{\frac{2\Delta}{\pi k_B T_\mathrm{bath}}}
          \sinh(\xi)\, K_0(\xi)
\end{equation}

\noindent with $\xi = hf_r/(2k_B T_\mathrm{bath})$ and $K_0$ the zeroth-order
modified Bessel function of the second kind.

The quasiparticle number evolves with bath temperature and read power
as~\cite{zmuidzinas2012superconducting, flanigan2018kinetic,
noroozian2012superconducting}

\begin{equation}
    N_\mathrm{qp} = \sqrt{\frac{V\,\Gamma_\mathrm{tot}}{\mathcal{R}}}
    \label{eq:N_qp}
\end{equation}

\begin{equation}
    \mathcal{R} = \frac{(2\times1.76)^3}{4\,N_0\,\Delta\,\tau_0\,f_\mathrm{trap}}
\end{equation}

\noindent where $\mathcal{R}$ is the effective quasiparticle recombination
constant; $\tau_0 = 290\,\mathrm{ns}$ is the electron-phonon interaction time
for thin-film Al, rescaled from Kaplan et~al.~\cite{kaplan1976quasiparticle}
using

\begin{equation}
  \tau_{0,\mathrm{thin}} = \tau_{0,\mathrm{thick}}
    \left(\frac{T_{c,\mathrm{thin}}}{T_{c,\mathrm{thick}}}\right)^{-3};
\end{equation}

\noindent and $f_\mathrm{trap}$ is a phonon trapping factor; for a thin
superconducting film on an insulating substrate, recombination phonons are
partially re-absorbed before escaping to the substrate, and
$f_\mathrm{trap} = 2$ (the standard approximation for this
geometry ~\cite{kaplan1976quasiparticle}). The total quasiparticle generation
rate is

\begin{align}
    \Gamma_\mathrm{tot}
        &= \Gamma_\mathrm{th} + \Gamma_\mathrm{pb}
         = \left(V\mathcal{R}\,n_\mathrm{th}^2\right)
         + \left(P_\mathrm{read}\,
           \frac{\eta_\mathrm{read}\chi_\mathrm{qp}}{\Delta}\right)
    \label{eq:Gamma}
\end{align}

\begin{equation}
    n_\mathrm{th} = 4 N_0 \Delta
        \sqrt{\frac{\pi k_B T_\mathrm{bath}}{2\Delta}}\,
        e^{-\Delta / k_B T_\mathrm{bath}}
\end{equation}

\noindent where $\eta_\mathrm{read}$ is the efficiency for converting
read-tone power into quasiparticles,
$\chi_\mathrm{qp} = Q_{i,\mathrm{qp}}^{-1}/Q_i^{-1}$ is the fraction of
internal dissipation due to quasiparticles, and $n_\mathrm{th}$ is the
thermal quasiparticle number density.

Using this model, $Q_{i,\mathrm{qp}}^{-1}$ has the temperature dependence
$Q_{i,\mathrm{qp}}^{-1} \propto e^{-1/T_\mathrm{bath}}\sinh(1/T_\mathrm{bath})
K_0(1/T_\mathrm{bath})$ and the read-power dependence
$Q_{i,\mathrm{qp}}^{-1} \propto \sqrt{1+P_\mathrm{read}}$, with dependence
on the film parameters $\alpha$, $\Delta$, and $\eta_\mathrm{read}$.

At high temperatures ($\geq 200\,\mathrm{mK}$) and intermediate read-tone
powers, $Q_{i,\mathrm{qp}}^{-1}$ strongly dominates total loss and
$\Gamma_\mathrm{th}$ dominates $N_\mathrm{qp}$. In this regime, for a fixed
read power, $Q_{i,\mathrm{tot}}^{-1}(T_\mathrm{bath})
\approx Q_{i,\mathrm{qp}}^{-1}(T_\mathrm{bath}) + C$ where $C$ is a small
constant offset, and contributions from $\Gamma_\mathrm{pb}$ can be
neglected. Fitting $Q_i^{-1}$ vs.\ $T_\mathrm{bath}$ data in this regime
yields $\alpha(\lambda/2,\,\lambda/4) = [0.26,\,0.26]$ and
$T_c(\lambda/2,\,\lambda/4) = [1.37,\,1.37]\,\mathrm{K}$
(Tables~\ref{tab:half_wave_params},~\ref{tab:quarter_wave_params}).
The fitted, resonator-specific $T_c$ values vary slightly from the 
feedline measurement of $T_c$ reported in
Table~\ref{tab:table1}; we use the resonator-derived values for all subsequent
analysis.

\begin{figure}[!t]
\centering
\includegraphics[width=\columnwidth]{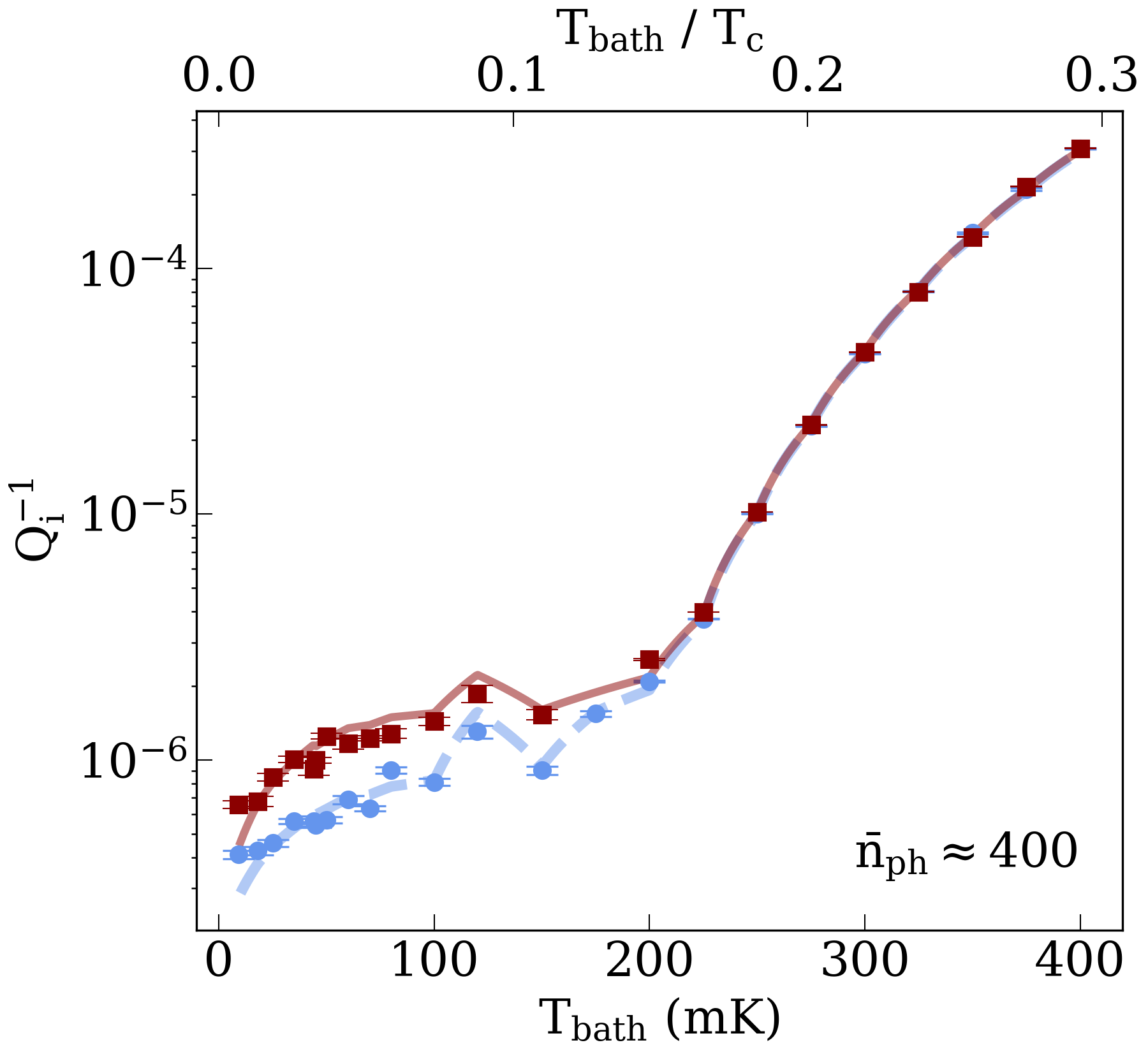}
\caption{Resonator loss $Q_i^{-1}$ vs.\ device temperature.
Red: $\lambda/2$ resonator data and fit.
Blue: $\lambda/4$ resonator data and fit.
Model values between data points result from interpolation.}
\label{fig:Qi_v_T}
\end{figure}

We then used these $\alpha$ and $T_c$ values to fit for $\eta_\mathrm{read}$.
At the highest read powers and lowest temperatures, $Q_{i,\mathrm{qp}}^{-1}$
dominates, so for a fixed low temperature $Q_{i,\mathrm{tot}}^{-1}(P_\mathrm{read})
\approx Q_{i,\mathrm{qp}}^{-1}(P_\mathrm{read}) + G$ with a constant offset $G$.
Using this method we find
$\eta_\mathrm{read}(\lambda/2,\,\lambda/4) = [1.7,\,1.8]\times10^{-3}$
(Tables~\ref{tab:half_wave_params},~\ref{tab:quarter_wave_params}).
This efficiency is small because the read tone lies far below the
pair-breaking threshold: at $f_r \approx 3.5$--$4\,\mathrm{GHz}$, the photon
energy $hf_r \approx 15$--$17\,\mu\mathrm{eV}$ is roughly $25\times$ smaller
than $2\Delta \approx 420\,\mu\mathrm{eV}$ for $T_c = 1.37\,\mathrm{K}$
aluminum, so single read photons cannot break Cooper pairs directly.
The nonzero $\eta_\mathrm{read}$ reflects pair-breaking via higher-order
phonon-mediated processes, which are inherently inefficient.

\subsubsection{TLS Loss}

Although the exact nature of parasitic two-level systems remains unclear, it
is thought that TLS arise when defects, impurities, or disorder appear in the
device substrate, the metal surface, or the substrate-metal interface. These
irregularities cause one or more atoms to tunnel between multiple available
energy states with a broad spectrum of transition energies. These fluctuating
sites can be excited by absorbing microwave photons from the resonator, and
eventually relax and re-emit phonons back into the resonator, producing
resonator transmission loss and phase noise.

TLS loss originates from surface and interface contamination, native surface
oxides, or bulk irregularities, with surface TLS often dominating. The
Standard Tunneling Model (STM) describes the aggregate TLS loss
as~\cite{gao2008physics}

\begin{equation}
    Q_{i,\mathrm{TLS}}^{-1}(\bar{n}_\mathrm{ph})
      = F\delta_\mathrm{TLS}^0\,\tanh(\xi)
        \left(1 + \frac{\bar{n}_\mathrm{ph}}{n_\mathrm{ph}^c}\right)^{-1/2}
    \label{eq:Q_TLS_STM}
\end{equation}

\noindent where $F$ is a geometric filling factor representing the fraction of
electric energy stored in TLS-hosting material, $\delta_\mathrm{TLS}^0$ is the
intrinsic TLS loss tangent at zero power and temperature, $\bar{n}_\mathrm{ph}$
is the average photon number in the device, and $n_\mathrm{ph}^c$ is the
critical photon number for full TLS saturation. The average photon number is
related to the read power by~\cite{mcrae2020materials}

\begin{equation}
    \bar{n}_\mathrm{ph} = \frac{P_\mathrm{read}}{2\pi m}
        \frac{Q_r^2}{Q_c} \frac{1}{h f_r^2}
    \label{eq:n_ph}
\end{equation}

\noindent where $m\,(\lambda/2,\,\lambda/4) = [1/2,\,1/4]$ is a geometric
factor for each resonator type.

This model describes resonator behavior under standard operating conditions,
but we observe it to deviate from our data at very low temperatures and photon
occupation numbers. The STM assumes a constant $n_\mathrm{ph}^c$; however,
it can be shown that $n_\mathrm{ph}^c \propto (T_1 T_2)^{-1}$, where the
average relaxation time $T_1$ is the timescale for a TLS to transition from
the excited to the ground state, and the average dephasing time $T_2$ is the
timescale over which phase coherence is preserved~\cite{phillips1987two}.
Both timescales are temperature-dependent; the relaxation timescale generally
follows $T_1 \propto \tanh(\xi)$~\cite{muller2015interacting}.
For a driven two-level system, the saturation condition requires the
stimulated transition rate to exceed both the relaxation rate $T_1^{-1}$ and
the dephasing rate $T_2^{-1}$, giving $n_\mathrm{ph}^c \propto
(T_1 T_2)^{-1}$~\cite{burnett2014evidence}.

Some works have found that TLS-TLS interactions are not always negligible,
with state changes in one TLS dephasing its neighbors~\cite{crowley2023disentangling}.
In this modified picture, as temperature decreases and more TLS revert to
their ground states, $T_2$ increases because fewer neighboring TLS state
changes occur. This relation has been modeled as $1/T_2 \propto
D\,T_\mathrm{bath}^\mu$, where $D$ is an empirically derived constant and
$\mu$ is a microscopic parameter relating to the TLS density of states
$\rho(E) \sim \rho_0(E/E_\mathrm{max})^\mu$~\cite{burnett2014evidence}. The
STM assumes negligible TLS-TLS interactions and $\mu = 0$, but experimental
works more commonly find $\mu \approx 1.1$--$2.0$~\cite{faoro2015interacting,
burnett2014evidence, crowley2023disentangling}. We therefore write

\begin{equation}
    n_\mathrm{ph}^c(T_\mathrm{bath}) \propto (T_1 T_2)^{-1}
        \propto \frac{D\,T_\mathrm{bath}^\mu}{\tanh(\xi)}
    \label{eq:n_ph_c}
\end{equation}

Then, following Ref.~\cite{crowley2023disentangling} we 
introduce an exponent $\beta$ such that
$\bar{n}_\mathrm{ph} \rightarrow \bar{n}_\mathrm{ph}^\beta$. 
The exponent $\beta$ is an empirical parameter introduced 
to account for the observation that different photon-mode 
occupancies saturate the TLS ensemble differently,
deviating from the uniform-saturation assumption of the STM\@. 

Combining these modifications and substituting~(\ref{eq:n_ph_c})
into~(\ref{eq:Q_TLS_STM}), we obtain the modified TLS loss model

\begin{equation}
    Q_{i,\mathrm{TLS}}^{-1}
      = F\delta_\mathrm{TLS}^0\,\tanh(\xi)
        \left(1 + \frac{\tanh(\xi)\,\bar{n}_\mathrm{ph}^\beta}
                       {D\,T_\mathrm{bath}^\mu}
        \right)^{-1/2}
    \label{eq:Q_TLS_modified}
\end{equation}

In the limiting case $\mu \to 0$, $\beta = 1$,
and $D \to n_\mathrm{ph}^c\,\tanh(\xi)$, our complex model 
(\ref{eq:Q_TLS_modified}) approaches the STM ~(\ref{eq:Q_TLS_STM}).

We first performed fits to our $Q_{i,\mathrm{tot}}^{-1}(T_\mathrm{bath})$
data at the lowest $P_\mathrm{read}$ (where TLS effects are most dominant)
using (\ref{eq:total_loss}), (\ref{eq:Q_qp}), and~(\ref{eq:Q_TLS_modified}),
with $\alpha$ and $T_c$ fixed from Section~II-B and with $\eta_\mathrm{read}$
and $Q_{i,\mathrm{other}}^{-1}$ as free parameters. This yields
$F\delta_\mathrm{TLS}^0(\lambda/2,\,\lambda/4)
= [3.87,\,3.47]\times10^{-6}$,
$D(\lambda/2,\,\lambda/4) = [386,\,492]$, and
$\mu(\lambda/2,\,\lambda/4) = [1.32,\,1.23]$
(Tables~\ref{tab:half_wave_params},~\ref{tab:quarter_wave_params}).

We then fixed $D$ and $\mu$ and fit our
$Q_{i,\mathrm{tot}}^{-1}(\bar{n}_\mathrm{ph})$ data
(Fig.~\ref{fig:Qi_vs_power}) across sixteen temperatures between
$9$--$400\,\mathrm{mK}$ to recover $F\delta_\mathrm{TLS}^0$, $\beta$, and
$\eta_\mathrm{read}$. This yields
$F\delta_\mathrm{TLS}^0(\lambda/2,\,\lambda/4)
= [3.85,\,3.29]\times10^{-6}$ and
$\beta(\lambda/2,\,\lambda/4) = [0.76,\,0.78]$
(Tables~\ref{tab:half_wave_params},~\ref{tab:quarter_wave_params}).

\subsubsection{Other Loss}

The final loss term, $Q_{i,\mathrm{other}}^{-1}$, is treated as a constant
representing the sum of all temperature- and power-independent loss
mechanisms (e.g., radiation loss, trapped magnetic vortex loss, and
stray-light-induced pair breaking). Our fits yield
$Q_{i,\mathrm{other}}^{-1}(\lambda/2,\,\lambda/4)
= [3.83,\,7.23]\times10^{-8}$
(Tables~\ref{tab:half_wave_params},~\ref{tab:quarter_wave_params}).

\subsection{Resonant Frequency Shifts}

We used the temperature-dependent shift in resonance frequency to
independently cross-check select material parameter fit results. The total
relative frequency shift is~\cite{gao2008experimental, kumar2008temperature}

\begin{equation}
    \delta f_\mathrm{tot}(T_\mathrm{bath})
        = \delta f_\mathrm{th}(T_\mathrm{bath})
        + \delta f_\mathrm{TLS}(T_\mathrm{bath})
    \label{eq:fr_shift}
\end{equation}

\noindent where $\delta f_\mathrm{tot}
= [f_r(T_\mathrm{bath}) - f_r(0)] / f_r(0)$, $\delta f_\mathrm{th}$ is the
thermal contribution, and $\delta f_\mathrm{TLS}$ is the TLS contribution.

When the bath temperature is high enough to produce pair-breaking phonons,
the quasiparticle population shifts the resonant frequency. This thermal
contribution is modeled as~\cite{gao2008experimental, kumar2008temperature}

\begin{equation}
    \delta f_\mathrm{th}
      = \frac{f_r(T_\mathrm{bath}) - f_r(0)}{f_r(0)}
      = -3.6\,\frac{\alpha\pi}{4}
        \sqrt{\frac{\pi k_B T_\mathrm{bath}}{2\Delta_0}}\,
        e^{-\Delta_0/k_B T_\mathrm{bath}}
    \label{eq:df_th}
\end{equation}
The coefficient 3.6 arises from the low-temperature limit of the
Mattis-Bardeen $S_2$ response function in the BCS model~\cite{zmuidzinas2012superconducting}.

TLS induce a change in $f_r$ by generating quasiparticles upon de-excitation
and through shifts in the real part of the dielectric constant. The TLS
frequency shift is modeled as~\cite{gao2008experimental, kumar2008temperature}

\begin{equation}
    \delta f_\mathrm{TLS}
      = \frac{F\delta_\mathrm{TLS}^0}{\pi}
        \left[\mathrm{Re}\!\left[\psi\!\left(\tfrac{1}{2}
              + \tfrac{\xi}{i\pi}\right)\right]
              - \ln\!\left(\tfrac{2\xi}{\pi}\right)\right]
    \label{eq:df_TLS}
\end{equation}

\noindent where $\psi$ is the digamma function.

At high temperatures ($\geq 200\,\mathrm{mK}$), $\delta f_\mathrm{th}$
dominates, allowing us to fit $f_r(T_\mathrm{bath})$ data to
(\ref{eq:df_th}) at fixed intermediate power and recover $\alpha = 0.25$
and $T_c = 1.37\,\mathrm{K}$. We then use these values with
(\ref{eq:fr_shift})--(\ref{eq:df_TLS}) to recover $F\delta_\mathrm{TLS}^0$
and $f_r(0)$ from a fit to the low-temperature $f_r(T_\mathrm{bath})$ data
at low power.

This yields $F\delta_\mathrm{TLS}^0(\lambda/2,\,\lambda/4)
= [3.71,\,3.10]\times10^{-6}$
(Tables~\ref{tab:half_wave_params},~\ref{tab:quarter_wave_params}).

\begin{figure}[!t]
\centering
\includegraphics[width=\columnwidth]{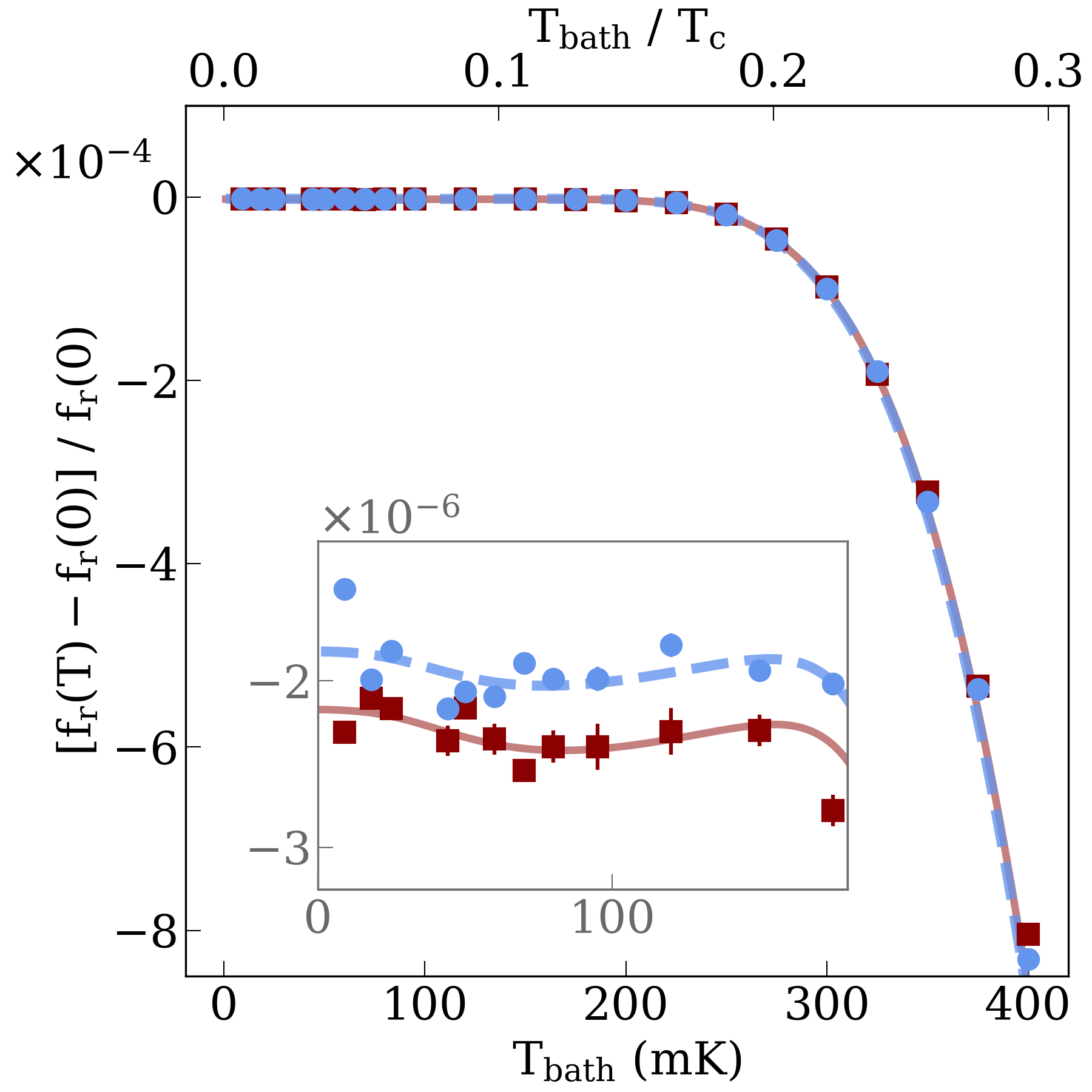}
\caption{Relative shift in resonant frequency vs.\ device temperature at low read power.
Red: $\lambda/2$ resonator. Blue: $\lambda/4$ resonator. Inset: magnified
view of the $0$--$175\,\mathrm{mK}$ regime where TLS effects dominate.}
\label{fig:df_vs_T}
\end{figure}

\section{Results and Discussion}

\noindent All fits were performed using bootstrapped least-squares fitting.
$Q_c$ is generally invariant across measurement conditions,
$Q_c(\lambda/2,\,\lambda/4) = [0.87,\,2.58]\times10^5$, as expected
(excluding the highest read-tone powers where the resonator response
bifurcates). We measured minimum
$Q_i^{-1}(\lambda/2,\,\lambda/4) = [3.64,\,8.57]\times10^{-8}$.
We note that these minima were achieved under different conditions
($\lambda/2$: $T_\mathrm{bath} = 60\,\mathrm{mK}$,
$\bar{n}_\mathrm{ph} \approx 4.8\times10^7$;
$\lambda/4$: $T_\mathrm{bath} = 18\,\mathrm{mK}$,
$\bar{n}_\mathrm{ph} \approx 5.6\times10^6$;
see Table~\ref{tab:table1}) and are not directly 
comparable as same-condition performance metrics.

The total loss in bulk superconducting aluminum has been measured with lower
bounds $\delta \approx (2$--$30)\times10^{-4}$~\cite{mcrae2020materials}.
For thin-film Al, $F\delta_\mathrm{TLS}^0$ varies widely with resonator
design and fabrication, spanning from a few $\times10^{-7}$~\cite{chayanun2024characterization,
richardson2016fabrication} to $1\times10^{-5}$~\cite{o2008microwave},
with typical values in the few $\times10^{-6}$
range~\cite{mcrae2020materials}. This variance may be attributed to
differing TLS populations, different resonator geometries (i.e.\ different
$F$ values), or differences in fitting models.

For our device we approximate $F \sim 8.1\times10^{-3}$ by rescaling the
Al$_2$O$_3$ calculations of Gao et~al.\ to our CPW
geometry~\cite{gao2008experimental}. Using this $F$ we estimate
$\delta_\mathrm{TLS}^0 \approx (3$--$5)\times10^{-4}$ for our thin-film Al.
For comparison, 480\,nm-thick AlO$_x$-covered Nb resonators yielded
$\delta_\mathrm{TLS}^0 \approx 1.3\times10^{-3}$~\cite{pappas2011two}.

\begin{table}[!t]
\caption{Film parameter fit results for the $\lambda/2$ resonator.
Column~(a): fits to $Q_i^{-1}(T_\mathrm{bath})$ at lowest $P_\mathrm{read}$.
Column~(b): simultaneous fits to $Q_i^{-1}(\bar{n}_\mathrm{ph})$ across all
temperatures. Column~(c): fits to $f_r(T_\mathrm{bath})$ at low
$P_\mathrm{read}$.}
\label{tab:half_wave_params}
\begin{center}
\renewcommand{\arraystretch}{1.2}
\begin{tabular}{lccc}
\hline
\textbf{Parameter} & \textbf{(a)} & \textbf{(b)} & \textbf{(c)} \\
\hline
$\alpha$
  & $0.255\pm0.006$ & ---            & $0.251\pm0.003$ \\
$T_c$ (K)
  & $1.367\pm0.003$ & ---            & $1.369\pm0.004$ \\
$F\delta_\mathrm{TLS}^0\,(\times10^{-6})$
  & $3.87\pm0.21$   & $3.85\pm0.23$  & $3.71\pm0.89$   \\
$D$$^\mathrm{a}$
  & $386\pm43$      & $364\pm54$     & ---             \\
$\mu$
  & $1.32\pm0.17$   & ---            & ---             \\
$\beta$
  & ---             & $0.76\pm0.05$  & ---             \\
$\eta_\mathrm{read}\,(\times10^{-3})$
  & ---             & $1.7\pm0.4$    & ---             \\
$Q_{i,\mathrm{other}}^{-1}\,(\times10^{-8})$
  & ---             & $3.83\pm1.3$   & ---             \\
\hline
\multicolumn{4}{l}{$^\mathrm{a}$Proportionality constant in (\ref{eq:n_ph_c});
see Section~III-B.}\\
\hline
\end{tabular}
\end{center}
\end{table}

\begin{table}[!t]
\caption{Film parameter fit results for the $\lambda/4$ resonator.
Columns as in Table~\ref{tab:half_wave_params}.}
\label{tab:quarter_wave_params}
\begin{center}
\renewcommand{\arraystretch}{1.2}
\begin{tabular}{lccc}
\hline
\textbf{Parameter} & \textbf{(a)} & \textbf{(b)} & \textbf{(c)} \\
\hline
$\alpha$
  & $0.259\pm0.01$  & ---            & $0.251\pm0.003$ \\
$T_c$ (K)
  & $1.366\pm0.007$ & ---            & $1.365\pm0.004$ \\
$F\delta_\mathrm{TLS}^0\,(\times10^{-6})$
  & $3.47\pm0.08$   & $3.29\pm0.40$  & $3.10\pm0.79$   \\
$D$$^\mathrm{a}$
  & $492\pm91$      & $441\pm76$     & ---             \\
$\mu$
  & $1.23\pm0.05$   & ---            & ---             \\
$\beta$
  & ---             & $0.78\pm0.04$  & ---             \\
$\eta_\mathrm{read}\,(\times10^{-3})$
  & ---             & $1.8\pm0.5$    & ---             \\
$Q_{i,\mathrm{other}}^{-1}\,(\times10^{-8})$
  & ---             & $7.23\pm0.7$   & ---             \\
\hline
\multicolumn{4}{l}{$^\mathrm{a}$Proportionality constant in (\ref{eq:n_ph_c});
see Section~III-B.}\\
\hline
\end{tabular}
\end{center}
\end{table}

We observed that at the lowest temperatures and photon occupation numbers
($T_\mathrm{bath} \leq 60\,\mathrm{mK}$, corresponding to
$hf_r/k_BT \gtrsim 3$), loss data deviated significantly from the STM
prediction. At these conditions the STM predicts TLS loss leveling off,
whereas our data show TLS loss continuing to decrease as temperature and
power approach zero.

Some works have attributed this anomalous behavior to TLS response bandwidths
narrowing with decreasing temperature, eventually excluding TLS with long
fluctuation timescales~\cite{tai2024anomalous}. We explored this approach
but achieved better fits with fewer free parameters using the modified model
presented in Section~III-B.

\begin{figure}[!t]
\centering
\includegraphics[width=\columnwidth]{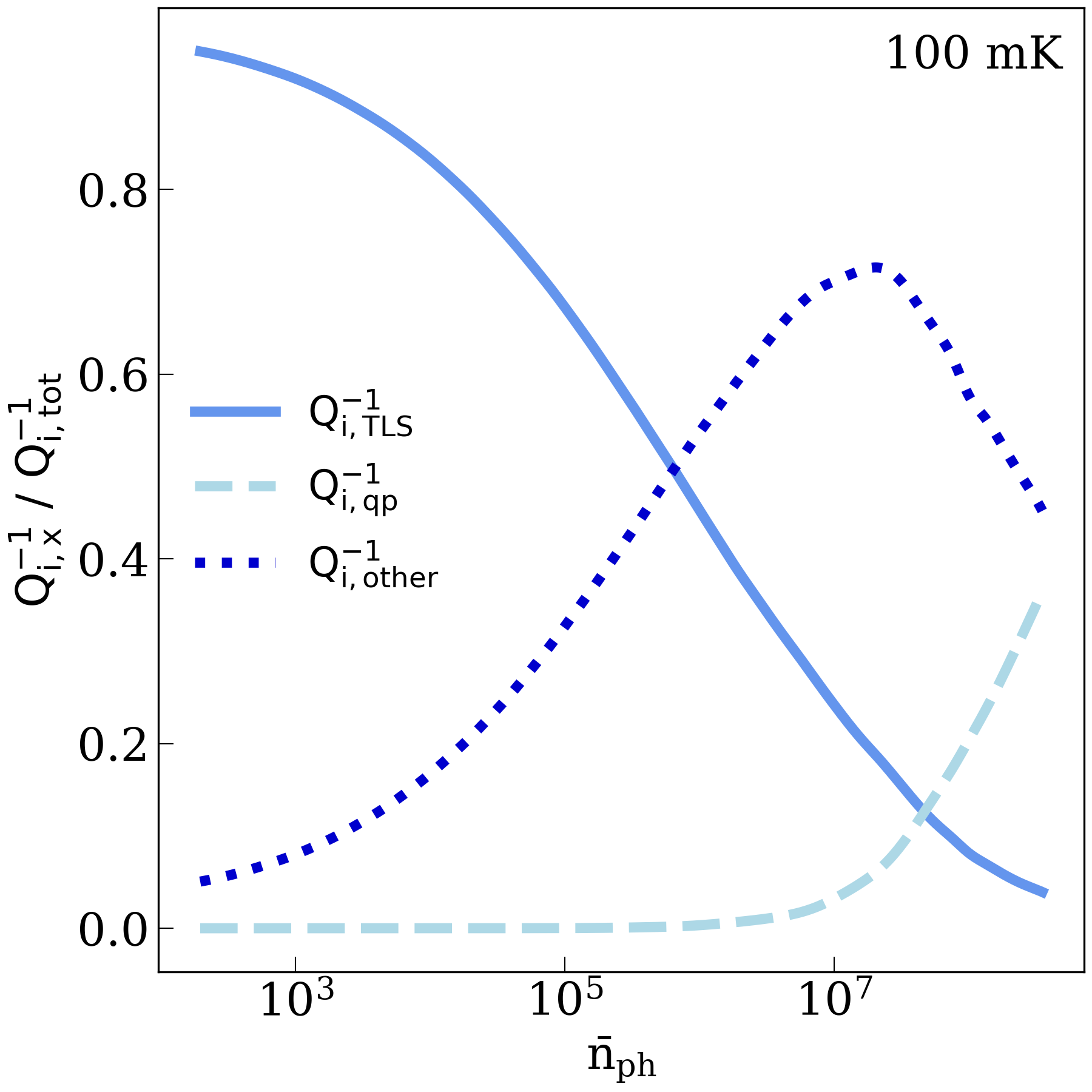}
\caption{Fractional contributions to total loss vs.\ average photon
occupation number $\bar{n}_\mathrm{ph}$ for the $\lambda/4$ resonator at
$100\,\mathrm{mK}$. Lines are derived from the model of Section~III with
interpolation. At high $\bar{n}_\mathrm{ph}$ the device enters a regime
dominated by the temperature- and power-independent intrinsic loss
$Q_{i,\mathrm{other}}^{-1}$.}
\label{fig:Qi_vs_power}
\end{figure}

Minimizing resonator volume increases the kinetic inductance fraction and
raises responsivity, but reduces the power-handling capability before
bifurcation occurs, locking the device out of the regime of optimal TLS
suppression. We designed our resonators to be antenna-coupled, which imposes
fewer surface-area constraints on the detector than other optical coupling
schemes and allows more freedom in resonator geometry. This enabled us to
design for high power-handling by keeping resonators wide and long, while
retaining a high kinetic inductance fraction through a very thin film.

Our wider CPW geometry dilutes the electric-field participation in lossy
dielectrics, since CPW fringe fields couple TLS losses in surface oxides and
the substrate to the resonator. The current density scales with conductor
surface area, so our CPW can accommodate higher powers before bifurcation.
This enabled an enhanced level of TLS suppression: at temperatures below
$200\,\mathrm{mK}$ we were able to increase $P_\mathrm{read}$ past the TLS
saturation point without bifurcating or entering a quasiparticle-dominated
regime. Instead, we observe a parameter space where temperature- and
power-independent intrinsic loss $Q_{i,\mathrm{other}}^{-1}$ dominates
(Fig.~\ref{fig:Qi_vs_power}), while TLS and quasiparticle loss each
contribute only $2$--$15\%$ of the total loss.

At the lowest temperatures this $Q_{i,\mathrm{other}}^{-1}$-dominated regime
spans a wide range of powers, narrowing and eventually disappearing as
$T_\mathrm{bath}$ increases due to thermal contributions to
$Q_{i,\mathrm{qp}}^{-1}$. This regime may be of particular interest for
studying intrinsic loss processes in detail. We posit that for devices where
TLS suppression is a priority, the design optimum should account for both
target responsivity and power-handling resonator volume simultaneously.

\section{Conclusion}

\noindent We have characterized diagnostic thin-film Al/Si CPW resonators
and performed an analysis of their dominant loss mechanisms. We find
$F\delta_\mathrm{TLS}^0 \sim\mathrm{few}\times10^{-6}$, on the lower end
of the typical range for this type of thin-film Al device.
At very low temperatures and photon occupation numbers
we observe deviations from the standard TLS loss model, and successfully
describe this regime with a modified model that accounts for
temperature-dependent TLS relaxation and dephasing times ($T_1$ and $T_2$)
and the differing saturation of TLS modes at varying photon occupancies
(via $\bar{n}_\mathrm{ph}^\beta$), recovering $\mu \approx 1.2$--$1.3$
consistent with values reported for interacting TLS populations in other
superconducting circuits.

While our $F\delta_\mathrm{TLS}^0$ is comparable to other thin-film Al
measurements, we observe a higher than typical level of TLS loss suppression
at high photon occupation numbers. We identify a regime at low temperatures
and intermediate read powers where inherent device loss
$Q_{i,\mathrm{other}}^{-1} \sim \mathrm{few}\times10^{-8}$ dominates over
both TLS and quasiparticle loss. We attribute this to our wide-feature CPW
design, which accommodates higher read powers before bifurcation and thereby
enables access to an optimally TLS-suppressed operating regime.

\section*{Acknowledgment}

The authors thank Mona Mirzaei for her role in device fabrication, and 
Andrew Harris, Serena Cronin, and Kyle Helson for project feedback and 
commentary. The devices were fabricated in the Detector Development 
Laboratory at NASA Goddard Space Flight Center.

\bibliography{aipsamp}

\begin{thebibliography}{10}
\providecommand{\url}[1]{#1}
\csname url@samestyle\endcsname
\providecommand{\newblock}{\relax}
\providecommand{\bibinfo}[2]{#2}
\providecommand{\BIBentrySTDinterwordspacing}{\spaceskip=0pt\relax}
\providecommand{\BIBentryALTinterwordstretchfactor}{4}
\providecommand{\BIBentryALTinterwordspacing}{\spaceskip=\fontdimen2\font plus
\BIBentryALTinterwordstretchfactor\fontdimen3\font minus
  \fontdimen4\font\relax}
\providecommand{\BIBforeignlanguage}[2]{{%
\expandafter\ifx\csname l@#1\endcsname\relax
\typeout{** WARNING: IEEEtran.bst: No hyphenation pattern has been}%
\typeout{** loaded for the language `#1'. Using the pattern for}%
\typeout{** the default language instead.}%
\else
\language=\csname l@#1\endcsname
\fi
#2}}
\providecommand{\BIBdecl}{\relax}
\BIBdecl

\bibitem{mcrae2020materials}
C.~R.~H. McRae, H.~Wang, J.~Gao, M.~R. Vissers, T.~Brecht, A.~Dunsworth, D.~P.
  Pappas, and J.~Mutus, ``Materials loss measurements using superconducting
  microwave resonators,'' \emph{Review of Scientific Instruments}, vol.~91,
  no.~9, 2020.

\bibitem{mirzaei2020mu}
M.~Mirzaei, E.~M. Barrentine, B.~T. Bulcha, G.~Cataldo, J.~A. Connors,
  N.~Ehsan, T.~M. Essinger-Hileman, L.~A. Hess, J.~W. Mugge-Durum, O.~Noroozian
  \emph{et~al.}, ``$\mu$-spec spectrometers for the exclaim instrument,'' in
  \emph{Millimeter, Submillimeter, and Far-Infrared Detectors and
  Instrumentation for Astronomy X}, vol. 11453.\hskip 1em plus 0.5em minus
  0.4em\relax SPIE, 2020, pp. 128--139.

\bibitem{wollack2014impedance}
E.~Wollack, D.~Chuss, K.~Rostem, and K.~U-Yen, ``Impedance matched absorptive
  thermal blocking filters,'' \emph{Review of Scientific Instruments}, vol.~85,
  no.~3, 2014.

\bibitem{khalil2012analysis}
M.~S. Khalil, M.~Stoutimore, F.~Wellstood, and K.~Osborn, ``An analysis method
  for asymmetric resonator transmission applied to superconducting devices,''
  \emph{Journal of Applied Physics}, vol. 111, no.~5, 2012.

\bibitem{gao2008physics}
J.~Gao, \emph{The physics of superconducting microwave resonators}.\hskip 1em
  plus 0.5em minus 0.4em\relax California Institute of Technology, 2008.

\bibitem{probst2015efficient}
S.~Probst, F.~Song, P.~A. Bushev, A.~V. Ustinov, and M.~Weides, ``Efficient and
  robust analysis of complex scattering data under noise in microwave
  resonators,'' \emph{Review of Scientific Instruments}, vol.~86, no.~2, 2015.

\bibitem{carter2016scraps}
F.~W. Carter, T.~S. Khaire, V.~Novosad, and C.~L. Chang, ``scraps: An
  open-source python-based analysis package for analyzing and plotting
  superconducting resonator data,'' \emph{IEEE Transactions on Applied
  Superconductivity}, vol.~27, no.~4, pp. 1--5, 2016.

\bibitem{gao2008experimental}
J.~Gao, M.~Daal, A.~Vayonakis, S.~Kumar, J.~Zmuidzinas, B.~Sadoulet, B.~A.
  Mazin, P.~K. Day, and H.~G. Leduc, ``Experimental evidence for a surface
  distribution of two-level systems in superconducting lithographed microwave
  resonators,'' \emph{Applied Physics Letters}, vol.~92, no.~15, 2008.

\bibitem{zmuidzinas2012superconducting}
J.~Zmuidzinas, ``Superconducting microresonators: Physics and applications,''
  \emph{Annu. Rev. Condens. Matter Phys.}, vol.~3, no.~1, pp. 169--214, 2012.

\bibitem{goldie2012non}
D.~Goldie and S.~Withington, ``Non-equilibrium superconductivity in
  quantum-sensing superconducting resonators,'' \emph{Superconductor Science
  and Technology}, vol.~26, no.~1, p. 015004, 2012.

\bibitem{flanigan2018kinetic}
D.~Flanigan, \emph{Kinetic inductance detectors for measuring the polarization
  of the cosmic microwave background}.\hskip 1em plus 0.5em minus 0.4em\relax
  Columbia University, 2018.

\bibitem{noroozian2012superconducting}
O.~Noroozian, \emph{Superconducting microwave resonator arrays for
  submillimeter/far-infrared imaging}.\hskip 1em plus 0.5em minus 0.4em\relax
  California Institute of Technology, 2012.

\bibitem{kaplan1976quasiparticle}
S.~B. Kaplan, C.~Chi, D.~Langenberg, J.-J. Chang, S.~Jafarey, and D.~Scalapino,
  ``Quasiparticle and phonon lifetimes in superconductors,'' \emph{Physical
  Review B}, vol.~14, no.~11, p. 4854, 1976.

\bibitem{phillips1987two}
W.~A. Phillips, ``Two-level states in glasses,'' \emph{Reports on Progress in
  Physics}, vol.~50, no.~12, p. 1657, 1987.

\bibitem{muller2015interacting}
C.~M{\"u}ller, J.~Lisenfeld, A.~Shnirman, and S.~Poletto, ``Interacting
  two-level defects as sources of fluctuating high-frequency noise in
  superconducting circuits,'' \emph{Physical Review B}, vol.~92, no.~3, p.
  035442, 2015.

\bibitem{burnett2014evidence}
J.~Burnett, L.~Faoro, I.~Wisby, V.~L. Gurtovoi, A.~V. Chernykh, G.~M.
  Mikhailov, V.~A. Tulin, R.~Shaikhaidarov, V.~Antonov, P.~J. Meeson
  \emph{et~al.}, ``Evidence for interacting two-level systems from the 1/f
  noise of a superconducting resonator,'' \emph{Nature Communications}, vol.~5,
  no.~1, p. 4119, 2014.

\bibitem{crowley2023disentangling}
K.~D. Crowley, R.~A. McLellan, A.~Dutta, N.~Shumiya, A.~P. Place, X.~H. Le,
  Y.~Gang, T.~Madhavan, M.~P. Bland, R.~Chang \emph{et~al.}, ``Disentangling
  losses in tantalum superconducting circuits,'' \emph{Physical Review X},
  vol.~13, no.~4, p. 041005, 2023.

\bibitem{faoro2015interacting}
L.~Faoro and L.~B. Ioffe, ``Interacting tunneling model for two-level systems
  in amorphous materials and its predictions for their dephasing and noise in
  superconducting microresonators,'' \emph{Physical Review B}, vol.~91, no.~1,
  p. 014201, 2015.

\bibitem{kumar2008temperature}
S.~Kumar, J.~Gao, J.~Zmuidzinas, B.~A. Mazin, H.~G. LeDuc, and P.~K. Day,
  ``Temperature dependence of the frequency and noise of superconducting
  coplanar waveguide resonators,'' \emph{Applied Physics Letters}, vol.~92,
  no.~12, 2008.

\bibitem{chayanun2024characterization}
L.~Chayanun, J.~Bizn{\'a}rov{\'a}, L.~Zeng, P.~Malmberg, A.~Nylander, A.~Osman,
  M.~Rommel, P.~L. Tam, E.~Olsson, P.~Delsing \emph{et~al.}, ``Characterization
  of process-related interfacial dielectric loss in aluminum-on-silicon by
  resonator microwave measurements, materials analysis, and imaging,''
  \emph{APL Quantum}, vol.~1, no.~2, 2024.

\bibitem{richardson2016fabrication}
C.~J. Richardson, N.~P. Siwak, J.~Hackley, Z.~K. Keane, J.~E. Robinson,
  B.~Arey, I.~Arslan, and B.~S. Palmer, ``Fabrication artifacts and parallel
  loss channels in metamorphic epitaxial aluminum superconducting resonators,''
  \emph{Superconductor Science and Technology}, vol.~29, no.~6, p. 064003,
  2016.

\bibitem{o2008microwave}
A.~D. O’Connell, M.~Ansmann, R.~C. Bialczak, M.~Hofheinz, N.~Katz, E.~Lucero,
  C.~McKenney, M.~Neeley, H.~Wang, E.~M. Weig \emph{et~al.}, ``Microwave
  dielectric loss at single photon energies and millikelvin temperatures,''
  \emph{Applied Physics Letters}, vol.~92, no.~11, 2008.

\bibitem{pappas2011two}
D.~P. Pappas, M.~R. Vissers, D.~S. Wisbey, J.~S. Kline, and J.~Gao, ``Two level
  system loss in superconducting microwave resonators,'' \emph{IEEE
  Transactions on Applied Superconductivity}, vol.~21, no.~3, pp. 871--874,
  2011.

\bibitem{tai2024anomalous}
T.~Tai, J.~Cai, and S.~M. Anlage, ``Anomalous loss reduction below two-level
  system saturation in aluminum superconducting resonators,'' \emph{Advanced
  Quantum Technologies}, vol.~7, no.~2, p. 2200145, 2024.

\end{thebibliography}
\bibliographystyle{IEEEtran}


\end{document}